\documentclass[prb,twocolumn,showpacs,preprintnumbers,amsmath,amssymb,superscriptaddress]{revtex4}

\usepackage{graphicx}
\usepackage{dcolumn}
\usepackage{bm}

\usepackage{amsmath}	
\begin{document}
\draft
\newcommand{\lw}[1]{\smash{\lower2.ex\hbox{#1}}}

\def\lesssim{\ \raise.3ex\hbox{$<$}\kern-0.8em\lower.7ex\hbox{$\sim$}\ }
\def\gesim{\ \raise.3ex\hbox{$>$}\kern-0.8em\lower.7ex\hbox{$\sim$}\ }

\title{Mott Phase in Polarized Two-component Atomic Fermi Lattice Gas:
A Playground for S=1/2 Heisenberg Model in Magnetic Field} 

\author{M.~Machida}
\email{machida.masahiko@jaea.go.jp}
\affiliation{CCSE, Japan Atomic Energy Agency, 6--9--3 Higashi-Ueno,
Taito-ku, Tokyo 110--0015, Japan}
\affiliation{CREST (JST), 4--1--8 Honcho, Kawaguchi, Saitama 332--0012,
Japan}
\author{M.~Okumura}
\email{okumura.masahiko@jaea.go.jp}
\affiliation{CCSE, Japan Atomic Energy Agency, 6--9--3 Higashi-Ueno,
Taito-ku, Tokyo 110--0015, Japan}
\affiliation{CREST (JST), 4--1--8 Honcho, Kawaguchi, Saitama 332--0012,
Japan}
\author{S.~Yamada} 
\email{yamada.susumu@jaea.go.jp}
\affiliation{CCSE, Japan Atomic Energy Agency, 6--9--3 Higashi-Ueno,
Taito-ku, Tokyo 110--0015, Japan}
\affiliation{CREST (JST), 4--1--8 Honcho, Kawaguchi, Saitama 332--0012, 
Japan}
\author{T.~Deguchi} 
\email{deguchi@phys.ocha.ac.jp}
\affiliation{Department of Physics, Graduate School of Humanities and
Sciences, Ochanomizu University, 2--1--1 Ohtsuka, Bunkyo-ku, Tokyo
112--8610, Japan} 
\author{Y.~Ohashi}
\email{yohashi@rk.phys.keio.ac.jp}
\affiliation{Faculty of Science and Technology, Keio University,
3--14--1, Hiyoshi, Kohoku-ku, Yokohama, Kanagawa 223--0061, Japan}
\affiliation{CREST (JST), 4--1--8 Honcho, Kawaguchi, Saitama 332--0012,
Japan}
\author{H.~Matsumoto}
\email{matumoto@ldp.phys.tohoku.ac.jp}
\affiliation{Institute for Materials Research, Tohoku University,
Katahira, Sendai 980--8577 Japan} 
\affiliation{Department of Physics, Tohoku University, Aramaki, Aoba, 
Sendai 980--8578 Japan} 
\affiliation{CREST (JST), 4--1--8 Honcho, Kawaguchi, Saitama 332--0012,
Japan}

\date{\today}

\begin{abstract} 
 We investigate effects of pseudo-spin population imbalance on Mott
 phases in 1D trapped two-component atomic Fermi gases loaded on optical
 lattices based on the repulsive Hubbard model in harmonic traps. By
 using the density matrix renormalization group method, we numerically
 calculate density profiles of each component and clarify the
 pseudo-spin magnetism. Consequently, we find that all the features from
 weakly imbalance to fully polarized cases are well described by $S=1/2$
 antiferromagnetic Heisenberg chain under magnetic field. These results
 indicate that the Mott phases offer experimental stages for studying
 various interacting spin systems. 
\end{abstract}

\maketitle

Recently, effects of population imbalance on interacting fermion systems
have been intensively studied in various fields as superconductors,
atomic Fermi gases, and quantum chromodynamics \cite{REVM}. The main
reason is recent drastic developments of experimental techniques in
superconductors and atomic Fermi gases \cite{ColdRev}. In particular, in
atomic Fermi gases, one can arbitrarily tune the population imbalance,
so that not only the so-called Fulde--Ferrell and Larkin--Ovchinikov
(FFLO) phase \cite{FFLO} with a spatially-modulated superfluid order
parameter but also the Chandrasekhar-Clogston limit \cite{CC} in a large
imbalance have been explored.

In cold atomic gases, besides the tunable imbalance, the optical lattice
and the variable interaction are like magic arts for condensed matter
physicists \cite{OLRev}. The optically-created periodical potential
flexibly builds up various playgrounds. The interaction tuning
associated with the Feshbach resonance provides a chance to
systematically study strongly-correlated behaviors \cite{OLRev}. In this
paper, we therefore study the population imbalance effect on the
strongly-correlated lattice stage, which is now one of the most
intensive but controversial issues in solid state matters \cite{CCI}. 

The atomic gas experiments usually employ the harmonic trap produced by
magnetic field and/or optical method to avoid the escape of atoms. The
harmonic trap brings about spatial inhomogeneities, which complicate the
observation of the quantum phase transition \cite{ExpIB}. Moreover, the
fact that the most convenient probe is atomic density profile have
limited the exploration of novel phases \cite{ExpIB}. For example, the
sign reversal in the FFLO superfluid order-parameter can not be directly
recognized by the density profile. Thus, the experimental confirmation
of FFLO still remains controversial in the trapped system
\cite{ExpIB,1DFFLO}. 

On the other hand, the Mott insulator core accompanied by metallic wings
predicted in the trapped optical lattice in the presence of the
repulsive interaction \cite{Rigol} can be easily confirmed by the
current probe like the density profile. These inhomogeneous phases have
been proposed by Quantum Monte Carlo studies \cite{Rigol} as well as the
exact diagonalization method \cite{Machida}. So far, theoretical studies
of the Mott core phase have been restricted to a particular case,
``balanced population''. In this paper, we focus on the Mott phase in 
the presence of population imbalance. Using the density matrix
renormalization group (DMRG) method \cite{White,DMRGRev}, we investigate
pseudo-spin structures by calculating density profiles of each component
in the Mott phase. Since the Mott core and its pseudo-spin structures
are directly observable, their exploration will be a suitable next
challenge in cold atom physics. 

Inside the Mott core, the on-site atomic density shows the unit-filling
and the density compressibility vanishes \cite{Rigol}. As a result, the
pseudo-spin degree of freedom solely survives, so that the core region
is well described by $S=1/2$ Heisenberg (local pseudo-spin interacting)
model for the two-component atomic Fermi gas. Moreover, we expect that a
population imbalance has a role of the magnetic field in the Heisenberg
model given by 
\begin{equation}
H_{\rm eff} = J \sum_{\langle i, j \rangle} {\bm S}_i \cdot {\bm S}_j -g 
\mu_B H_{\rm ext} \sum_i S_i^z \, , \label{Heff}
\end{equation}
where the fictitious magnetic field $H_{\rm ext}$ is varied by the
magnitude of the population imbalance in the original system. In this
paper, we explicitly confirm that the spin structure in the Mott core
region is really described by the effective Hamiltonian (\ref{Heff})
using the DMRG method \cite{White,DMRGRev}. Namely, we suggest that the
Mott core can be employed as a model system to widely study the
magnetism in interacting spin models. One of the advantages using the
equivalence is that one can easily reach a very high field
range. Moreover, although this paper concentrates on the one-dimensional
and two-component Fermi atom system as a trial problem, higher
dimensional, frustrated, and large $S$ cases are also possible to study. 

The starting model Hamiltonian \cite{Rigol,Machida} describing trapped
two-component Fermi atoms under the 1D strong optical lattice is given
by the 1D Hubbard model with the harmonic trap,
\begin{align}
H_{\rm Hubbard} & = -t \sum_{\langle i,j \rangle,\sigma} \left( c_{i
\sigma }^\dag c_{j \sigma } + {\rm H.c.} \right) + U \sum_i  n_{i
 \uparrow} n_{i \downarrow}  \nonumber \\  
& \quad {} + V \left( \frac{2}{N-1} \right)^2 \sum_{i,\sigma} \left( i -
 \frac{N+1}{2} \right)^2 n_{i \sigma} \, , \label{HHub}
\end{align}
where, the summation for the pseudo-spin $\sigma$ is taken over
two-components assigned as $\sigma=\uparrow$ and $\downarrow$,
respectively. $c_{i\sigma}^\dagger$ is the creation operator of a Fermi
atom with the pseudo-spin $\sigma$ at the $i$-th lattice state, and
$n_{i\sigma} (\equiv c_{i \sigma}^\dag c_{i \sigma})$ is the site
density one for the same pseudo-spin. In the first term of the
Hamiltonian (\ref{HHub}), $t$ describes the nearest-neighbor hopping
parameter and the summation $\langle i,j \rangle$ is taken over the
nearest-neighbor sites, and $U$ ($>0$) in the second term is the on-site
repulsive interaction. The last term in Eq.~(\ref{HHub}) describes a
harmonic trap potential, where $V$ is the potential height at the edge
sites. $N$ is the total number of lattice sites, and $N_{\rm F}$ is that
of fermions with $\sigma=\uparrow$ and $\downarrow$ ($N_{\rm F} \equiv
N_{\uparrow} + N_{\downarrow}$). Throughout this paper, an atom
component with $\sigma = \uparrow$ is always a major one. As a main
numerical method, we employ the DMRG to explore the ground state of the
model (\ref{HHub}). At first, the number of states kept ($m$) in DMRG is
selected by a comparison of the ground state energy with the exact
diagonalization method for small size ($N=20$). In larger sizes, we
select $m$ which gives no significant difference by increasing $m$
further. For $N=60(120)$ and $180(240)$ in the Hubbard model, we confirm
that $m=100$ and $m=300$ is enough, respectively. In addition, for
$N=60$ in the Heisenberg model, $m=100$ is selected due to the same
reason. 
 
\begin{figure}
\includegraphics[scale=0.37]{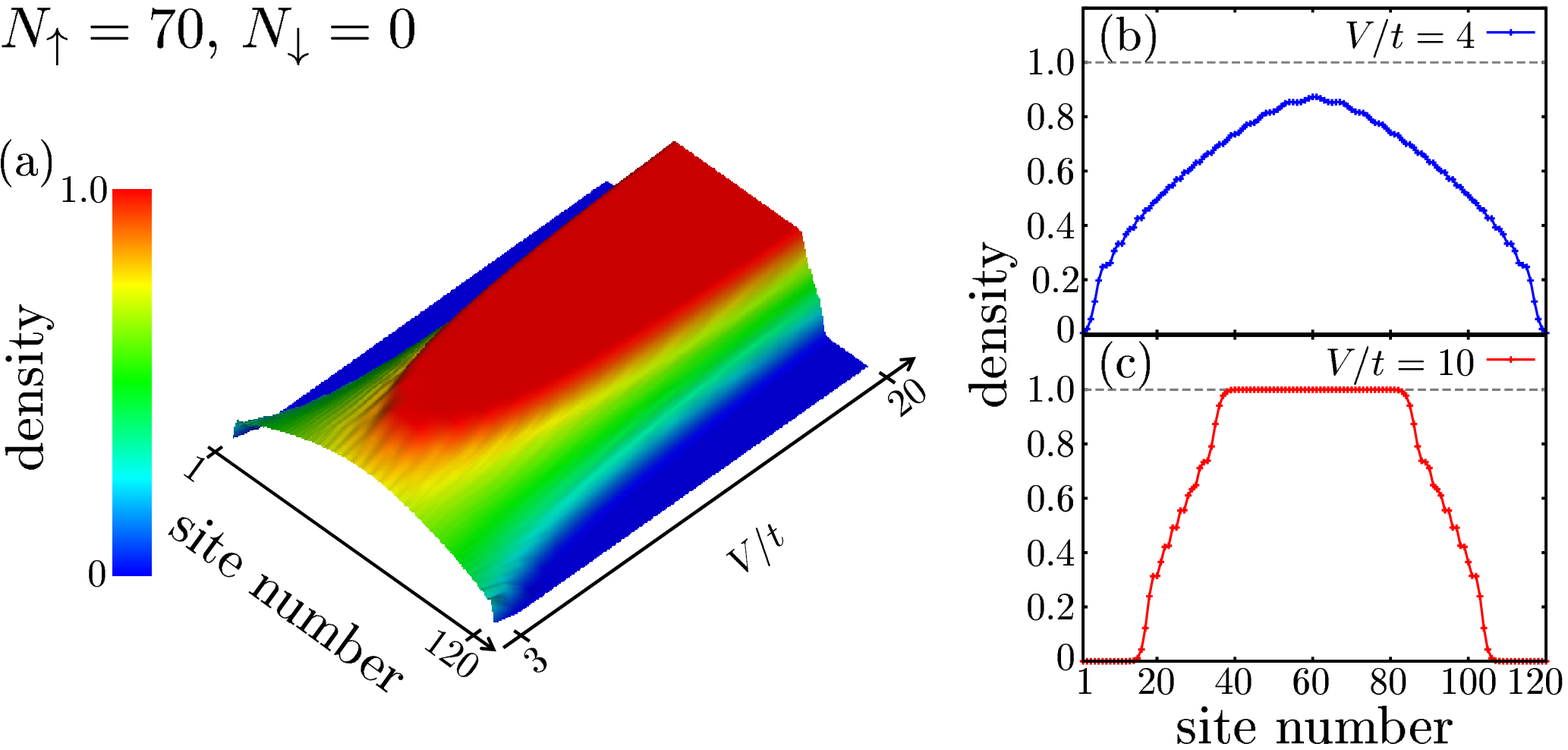}
\caption{\label{fig1} (a) The trap-potential strength $V$ dependence of
 the particle density profile $n_{\rm tot} (i) (= n_{\uparrow} (i))$
 for a completely polarized fermionic gas with slice pictures for two
 cases (b) $V/t = 4$ and (c) $10$.} 
\end{figure}

Let us show DMRG results of the model (\ref{HHub}). Firstly, we show
atomic density profiles in the case of the perfect polarization
($P\equiv (N_\uparrow-N_\downarrow)/N=1$) in Fig.~\ref{fig1}. When
$V/t\gesim 5$, we find the insulating core in the center of the trap,
over which the unit filling is spread. Since the compressibility is zero
and the polarization is perfect in this insulating core, it is regarded
as a ferromagnetic insulator. We note that this insulating state
originates from only the Pauli's exclusion principle and differs from
the Mott state caused by a repulsive interaction between fermions. We
also point out that this ferromagnetic insulating core can be described
by the antiferromagnetic Heisenberg model (\ref{Heff}) in the presence
of an infinitely strong magnetic field.

\begin{figure}
\includegraphics[scale=0.37]{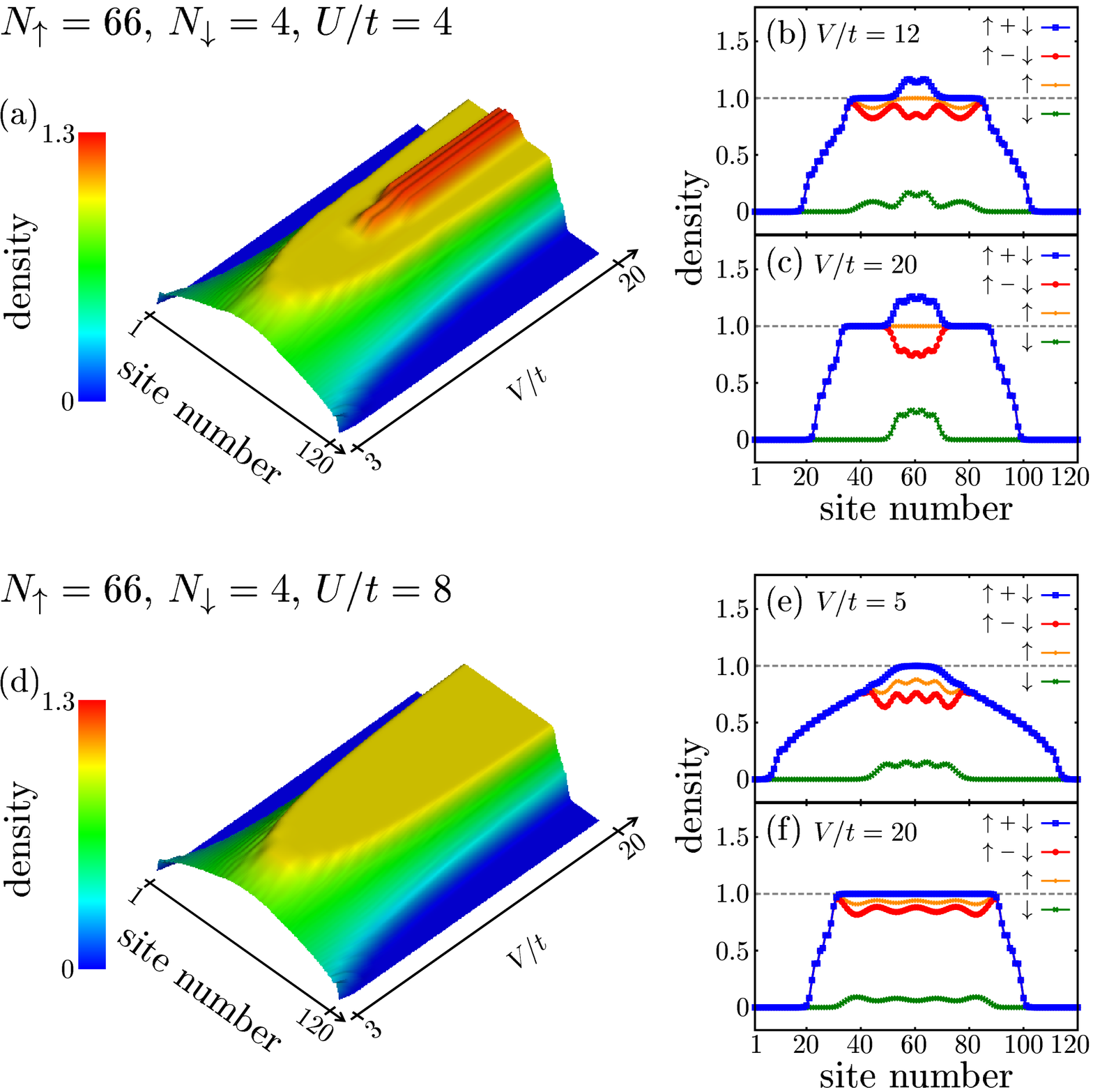}
\caption{\label{fig2} The trap-potential strength $V$ dependences of
 the particle density profiles $n_{\rm tot} (i) (= n_{\uparrow} (i) +
 n_{\downarrow} (i))$ for $N_{\uparrow} =66$ and $N_{\downarrow}=4$ in
 (a) $U/t=4$ with two slice pictures at (b) $V/t = 12$ and (c) $20$ and
 (d) $U/t=8$ with the same ones at (e) $V/t=12$ and (f) $20$.} 
\end{figure}

Next, let us study cases in which the minority spin component slightly
increases from the zero (the complete polarized one). We examine density
profiles in two typical situations, i.e., those in the presence of
relatively weak and strong repulsive interaction. The upper and the
lower panels in Fig.~\ref{fig2} are $V/t$ dependences of density
profiles of the former $(U/t=4)$ and the latter $(U/t=8)$ cases,
respectively. As seen in Fig.~\ref{fig2}, the unity core is broken in
the central region about above $V/t=10$ in the weak interaction case
(Fig.~\ref{fig2}(a)), while its flat plateau feature is still kept up to
$V/t=20$ (Fig.~\ref{fig2}(d)) in the strong interaction one. Here, we
note that the breakdown of the unity core is also observed above
$V/t=20$ in the strong interaction $(U/t =8)$ case. Namely, the
$V$-dependent changes in the density profiles are qualitatively
equivalent in both the cases. On the other hand, we find from these
results that the unity core is the so-called Mott state since its phase
stability actually depends on the interaction strength.

\begin{figure}
\includegraphics[scale=0.37]{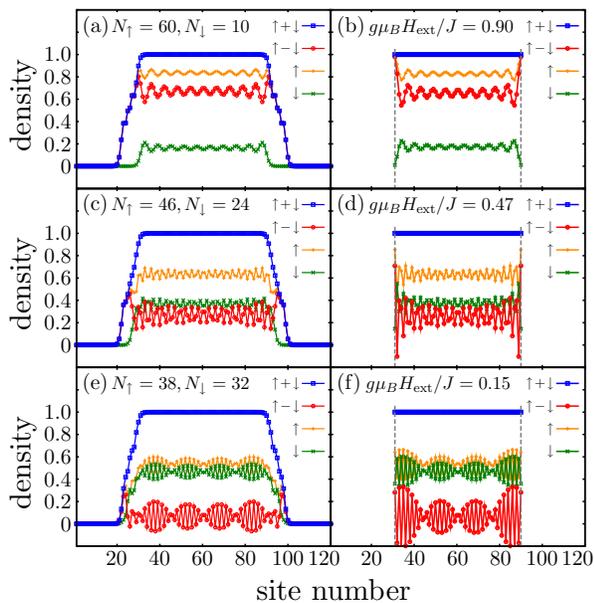}
\caption{\label{fig3} The profile changes in the Hubbard model
 (\ref{HHub}) with decreasing the population imbalance ratio at $U/t=8$ 
 and $V/t=20$. In the fixed total particle number $N_{\rm F}=70$, (a)
 $N_{\uparrow} = 60$, (c) $46$, and (e) $38$. The spin densities in the
 60-sites Heisenberg chain with the open boundary condition in the
 external magnetic field given by Eq.~(\ref{Heff}) are plotted with (b)
 $g \mu_B H_{\rm ext} / J = 0.90$, (d) $0.47$, (f) $0.15$.}
\end{figure}

Now, let us concentrate on pseudo-spin structures inside the Mott-phase 
as seen in Figs.~\ref{fig2}(e) and \ref{fig2}(f). Before the Mott phase
destruction occurs, we find that the minority makes a profile like
Wigner lattice inside the Mott core. The number of the peak in the
minority profile is the same as that of the minority atoms as seen in
Figs.~\ref{fig2}(e) and \ref{fig2}(f), where the number is just four
(see Fig.~\ref{fig3}(a) for another case in which the number is
ten). These Wigner lattice like profiles can be explained by the
antiferromagnetic Heisenberg model (\ref{Heff}) in finite but strong
magnetic field. The effective model (\ref{Heff}) then predicts the spin
density wave (SDW) state whose periodicity is characterized by $2k_F =
\pi (1-\bar{m})$ where $\bar{m}$ is the magnetization normalized by the
saturated magnetization and $k_F$ is the Fermi wave vector in the
equivalent spinless fermion system \cite{Giamarchi}. In the present
imbalance system, since the $\bar{m}$ is a controlable parameter via the
population imbalance, the periodicity is given by 
\begin{equation}
2k_F = \pi \left[ 1 - \left( \frac{N_{\uparrow}^{\rm Mott} -
N_{\downarrow}^{\rm Mott}}{N^{\rm Mott}} \right) \right] ,
\end{equation}
where $N_{\uparrow}^{\rm Mott}$ and $N_{\downarrow}^{\rm Mott}$ are the
numbers of the up- and down-spin particles participating the Mott core,
respectively, and $N^{Mott}$ is the number of the lattice sites occupied
by the Mott core. Thus, one finds why the minority profile shows Wigner
crystal like ones, e.g., $2k_F= \pi[1-(56-4)/60] = 2 \pi \cdot 4/60$ in 
Fig.~\ref{fig2}(f) and $2k_F = \pi [1-(50-10)/60] = 2\pi \cdot 10/60$ in
Fig.~\ref{fig3}(a), where the Mott phase covers $60$ sites ($N^{\rm
Mott}=60$), $N_{\uparrow}^{\rm Mott} = N_{\uparrow}-10$ ($10$ majority
particles contribute to make the metallic wings), and
$N_{\downarrow}^{\rm Mott}=N_{\downarrow}$ as seen in Fig.~\ref{fig3}(f)
and Fig.~\ref{fig3}(a). These profiles are really confirmed by the DMRG
calculation of 60-sites Heisenberg chain model in a magnetic field with
the open boundary condition e.g., compare Fig.~\ref{fig3}(a) with
Fig.~\ref{fig3}(b). This result clearly demonstrates that the imbalanced
Mott phases in the trapped Fermi lattice systems are equivalent with the
effective interacting spin model under magnetic field. 

\begin{figure}
\includegraphics[scale=0.6]{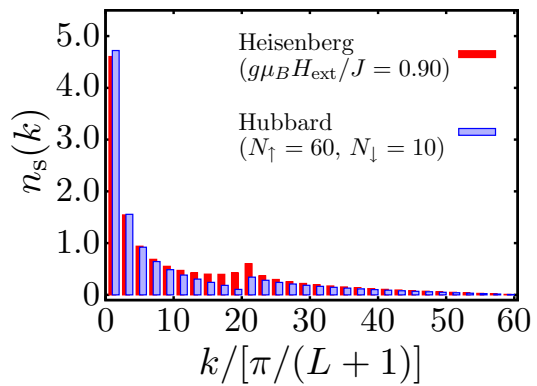}
\caption{\label{fig4} The comparison of $k$ vs. the Fourier transformed
 spin density $n_{\rm s} (k)$ for Figs.~\ref{fig3}(a) and \ref{fig3}(b),
 i.e., the trapped Hubbard model Eq.~(\ref{HHub}) and the Heisenberg
 model Eq.~(\ref{Heff}) with the open boundary condition, in which $k
 \equiv (\pi \ell)/(L+1)$ $(\ell=1,2,...)$, where $L =60$. In the case
 of the trapped Hubbard model, only the central core region is used for
 the Fourier transformation.} 
\end{figure}

Let us compare the spin density distributions of the Mott core with ones
of the Heisenberg model in more details. For the purpose, we evaluate
the Fourier component $n_{\rm s}(k)$ of the spin density distributions
$n_{\rm s} (i) (= n_{\uparrow} (i) - n_{\downarrow} (i))$ in a central
range (from $i=31$ to $90$, i.e., $L=60$) shown in
Figs.~\ref{fig3}(a)--\ref{fig3}(b). Figures \ref{fig4} show $k \equiv
(\pi \ell)/(L+1)$ $(\ell=1,2,\cdots,L)$ vs.~$n_{\rm s}(k)$. In these
figures, one can find that a main peak characterizing the SDW structure
(e.g., $\ell=21$) and other profiles almost coincide between both
cases. This result indicates that the Mott phases confined inside the
harmonic trap can be well described by the effective Heisenberg model
with the open boundary condition. 

\begin{figure}
\includegraphics[scale=0.37]{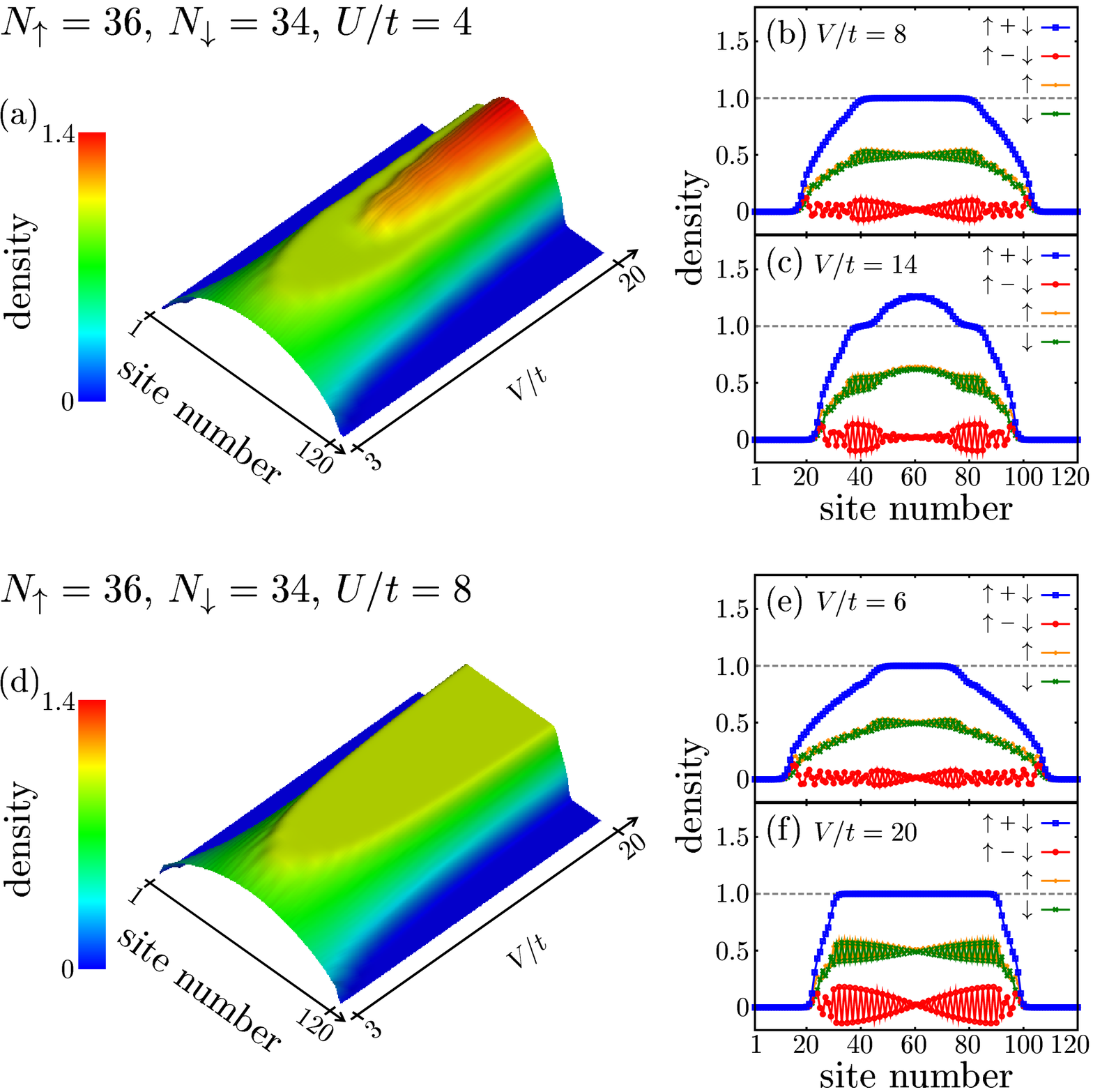}
\caption{\label{fig5} The trap potential strength $V$ dependences of
 profiles of the particle density $n_{\rm tot} (i) (= n_{\uparrow} (i) +
 n_{\downarrow} (i))$ for $N_{\uparrow} =36$ and $N_{\downarrow}=34$ in
 (a) $U/t=4$ with two slice pictures at (b) $V/t = 12$ and (c) $20$ and
 (d) $U/t=8$ with the same ones at (e) $V/t=12$ and (f) $20$.}  
\end{figure}

We further decrease the population imbalance $P$, i.e., increase the
number of the minority atoms. Then, the results, e.g.,
Fig.~\ref{fig3}(c) reveals that the SDW periodicity is reduced according
to $2k_F = \pi (1-\bar{m})$. We also note that by further imbalance
decrease, in addition to the SDW spin configuration, another modulation
structure with a wave length being much longer than the lattice constant
appears [see Fig.~\ref{fig3}(e)]. This is regarded to emerge as a
boundary effect since the incommensuration of $2k_F = \pi (1-\bar{m})$
with the lattice becomes visible, i.e., a beating modulation whose
periodicity given by $\pi \bar{m}$ is exposed. For example, $2k_F =
\pi[1-(33-27)/60]=\pi(1-6/60)$ in Fig.~\ref{fig3}(e), where it is noted
that both the majority (5 particles) and the minority (5 particles)
equally contribute to the metallic wing. See another case, $2k_F = \pi
[1-(31-29)/60] =\pi(1-2/60)$ in Fig.~\ref{fig5}(f), where 5 majority and
5 minority particles also participate the metallic wing similar to
Fig.~\ref{fig3}(e). As shown in Figs.~\ref{fig3}(d) and \ref{fig3}(f),
the change of the spin structure seen in Figs.~\ref{fig3}(c) and
\ref{fig3}(e) in the Hubbard model (\ref{HHub}) can be also well
reproduced by decreasing the strength of the magnetic field in the
effective model (\ref{Heff}). One finds that even the beating modulation 
due to the boundary effect is also reproduced. 

\begin{figure}
\vspace*{1mm}
\includegraphics[scale=0.37]{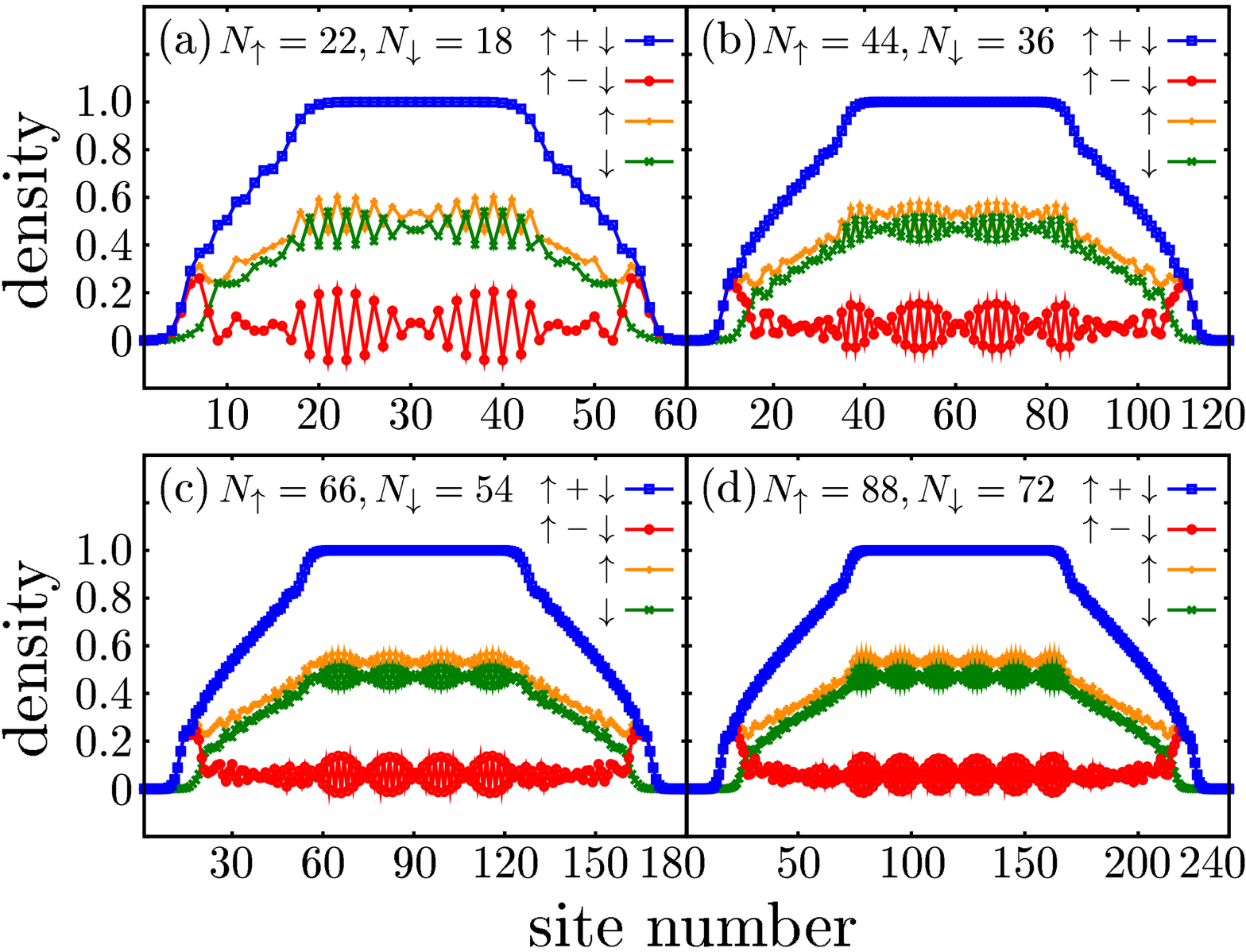}
\caption{\label{fig6} The site number $N$ and the total atom number
 $N_{\rm F}$ dependences of the atom profiles with keeping the imbalance 
 ratio for (a) $N_{\uparrow} = 22$ and $N= 60$, (b) $N_{\uparrow}=44$
 and $N=120$, (c) $N_{\uparrow} = 66$ and $N=180$, and (d) $N_{\uparrow}
 = 88$ and $N=240$. In these cases, $U/t = 20$, and $V/t = 6$.}
\end{figure}

Let us turn to further small imbalance cases close to the balanced one. 
The upper and lower panels of Fig.~\ref{fig5} show $V/t$ dependent
profiles in which the population ratio is $36:34$ in strong and weak
$U/t$, respectively. The profile in the weak interaction shows that the
Mott insulator core is broken about above $V/t = 12$ and the almost
antiferromagnetic staggered profile is lost in the broken region as
shown in Fig.~\ref{fig5}(c). The loss of the staggered structure is also
observed in the periphery \cite{Note1} around the Mott core as seen in
Fig.~\ref{fig5}(b) (see Fig.~\ref{fig6}(a) for another case). These
results clearly reflect that the staggered profile, i.e., the SDW phase
is formed only by the spin degree of freedom. The staggered profile
diminishes in the metallic region in which the charge degree of freedom
is alive. In addition, inside the Mott core, another long modulation is
also observed in both the weak and strong interaction cases as seen in
Figs.~\ref{fig5}(b) and \ref{fig5}(f). In order to check the size
dependence of this modulation, we examine the profiles by simply
increasing both the lattice sites and the number of total atoms with
keeping the population imbalance ratio a constant. The modulation and
its wave periodicity is found to be almost size-independent within the
range as seen in Figs.~\ref{fig6}(a)--\ref{fig6}(d). These results
indicate that such a modulation is clearly observable in 1-D atomic
Fermi gases loaded on optical lattices. In addition, we note that the
effective Heisenberg model with the open boundary condition can
reproduce these results.

We investigated the repulsively-interacting polarized 1-D Hubbard model
with harmonic confinement potentials by using the DMRG method. Inside
the core phase (where the site density equals to the unit) emerged
universally for arbitrary $P$, we found that its spin structure is
described by the antiferromagnetic Heisenberg model in magnetic
field. This equivalence was confirmed by DMRG calculations for both the
original and effective models. We suggest that the
repulsively-interacting polarized trapped lattice fermion systems offer
various playgrounds of not only the Hubbard type but also the
interacting localized-spin one. This idea may have a new impact on
studies of the magnetism in the solid state physics. 

One of authors (M.M.) thank T.~Koyama, M.~Kato, T.~Ishida, H.~Ebisawa, 
N.~Hayashi, and T.~Sakai for helpful discussions about
superconductivity. The work was partially supported by Grant-in-Aid for
Scientific Research on Priority Area ``Physics of new quantum phases in
superclean materials'' (Grant No.~20029019) from the Ministry of
Education, Culture, Sports, Science and Technology of Japan. This work
was also supported by Grant-in-Aid for Scientific Research from MEXT,
Japan (Grant No.~20500044). One of authors (M.M.) is supported by JSPS
Core-to-Core Program-Strategic Research Networks, ``Nanoscience and
Engineering in Superconductivity (NES)''. 



\begin{references}
\bibitem{REVM} R.~Casalbuoni and G.~Nardulli, Rev. Mod. Phys. {\bf
 76}, 263 (2004). 
\bibitem{ColdRev} See for reviews, e.g., I.~Bloch, J.~Dalibard, and
 W~Zwerger, Rev. Mod. Phys. {\bf 80}, 885 (2008); S.~Giorgini,
 L.~P.~Pitaevskii, and S.~Stringari, Rev. Mod. Phys. {\bf 80}, 1215
 (2008), and references therein.
\bibitem{FFLO} P.~Fulde and R.~A.~Ferrell, Phys. Rev. {\bf 135}
 A550 (1964); A.~I.~Larkin and Y.~N.~Ovchinnikov, Sov. Phys. JETP,  
{\bf 20}, 762 (1965). 
\bibitem{CC} Y.~Shin, C.~H.~Schunck, A.~Schirotzek, and W.~Ketterle, 
Nature {\bf 451}, 689 (2008). 
\bibitem{OLRev} For a review, see, e.g., M.~Lewenstein, A.~Sanpera,
 V.~Ahufinger, B.~Damski, A.~Sen(De), and U.~Sen, Adv. Phys. {\bf 56},
 243 (2007), and references therein.  
\bibitem{CCI} A.~Bianchi, R.~Movshovich, C.~Capan, P.~G.~Pagliuso, and
 J.~L.~Sarrao, Phys. Rev. Lett. {\bf 91}, 187004 (2003); K.~Kakuyanagi,
 M.~Saitoh, K.~Kumagai, S.~Takashima, M.~Nohara, H.~Takagi, and
 Y.~Matsuda, Phys. Rev. Lett. {\bf 94}, 047602 (2005); V.~F.~Correa,
 T.~P.~Murphy, C.~Martin, K.~M.~Purcell, E.~C.~Palm,
 G.~M.~Schmiedeshoff, J.~C.~Cooley, and S.~W.~Tozer, Phys. Rev. Lett. 
 {\bf 98}, 087001 (2007). 
\bibitem{ExpIB} For experiments, see e.g., M.~W.~Zwierlein,
 A.~Schirotzek, C.~H.~Schunck, and W.~Ketterle, Science {\bf 27}, 492
 (2006); G.~B.~Partridge, W.~Li, R.~I.~Kamar, Y.~Liao, and R.~G.~Hulet,
 {\it ibid}. {\bf 27}, 503 (2006); M.~W.~Zwierlein, C.~H.~Schunck,
 A.~Schirotzek, and W.~Ketterle, Nature {\bf 442}, 54 (2006); Y.~Shin,
 M.~W.~Zwierlein, C.~H.~Schunck, A.~Schirotzek, and W.~Ketterle,
 Phys. Rev. Lett. {\bf 97}, 030401 (2006); C.~H.~Schunck, Y.~Shin,
 A.~Schirotzek, M.~W.~Zwierlein, and W.~Ketterle, Science {\bf 11}, 867
 (2007). 
\bibitem{1DFFLO} For theoretical works, see e.g., A.~Moreo and
 D.~J.~Scalapino, Phys. Rev. Lett. {\bf 98}, 216402 (2007);
 A.~E.~Feiguin and F.~Heidrich-Meisner, Phys. Rev. B {\bf 76}, 220508(R)
 (2007); M.~Tezuka and M.~Ueda, Phys. Rev. Lett. {\bf 100}, 110403
 (2008); G.~G.~Batrouni, M.~H.~Huntley, V.~G.~Rousseau, and
 R.~T.~Scaletter, {\it ibid}. {\bf 100}, 116405 (2008); M.~Rizzi,
 M.~Polini, M.~A.~Cazalilla, M.~R.~Bakhtiari, M.~P.~Tosi, and R.~Fazio,
 Phys. Rev. B {\bf 77}, 245105 (2008); A.~L{\" u}scher,
 R.~M.~Noack, and A.~M.~L{\" a}uchli, Phys. Rev. A {\bf 78}, 013637
 (2008); M.~Casula, D.M.~Ceperley, and E.J.~Mueller, arXiv:0806.1747;
 A.E.~Feiguin and F.~Heidrich-Meisner, arXiv:0809.1539; A.E.~Feiguin and
 D.A.~Huse, arXiv:0809.3024. 
\bibitem{Rigol} M.~Rigol and A.~Muramatsu, Phys. Rev. A {\bf 69}, 053612 
 (2004). 
\bibitem{Machida} M.~Machida, S.~Yamada, Y.~Ohashi, and H.~Matsumoto,
 Phys. Rev. Lett. {\bf 93}, 200402 (2004). 
\bibitem{White} S.~R.~White, Phys. Rev. Lett. {\bf 69}, 2863 (1992);
 Phys. Rev. B {\bf 48}, 10345 (1993).
\bibitem{DMRGRev} For recent reviews, see e.g., U.~Schollw{\" o}ck,
 Rev. Mod. Phys. {\bf 77}, 259 (2005); K.~A.~Hallberg, Adv. Phys. {\bf
 55}, 477 (2006), and references therein.
\bibitem{Giamarchi} T.~Giamarchi, {\it Quantum Physics in One Dimension}
 (Oxford University Press, New York, 2004).
\bibitem{Note1} In the metallic peripheries in Fig.~\ref{fig5}(b),
 \ref{fig5}(c) and \ref{fig5}(e), antiferromagnetic like zigzag
 configurations are found, but the periodicities are larger than the
 lattice constant.
\end{references}
\end{document}